\documentclass[showpacs,aps,prl,groupedaddress,superscriptaddress,twocolumn]{revtex4}
\usepackage{amsthm}
\usepackage{amssymb}
\usepackage{amsmath}
\usepackage{graphicx}

\def\p0{\phi_{0}}

\def\vecr{{\bf r}}
\def\vechr{{\bf \hat r}}

\def\veck{{\bf k}}
\def\vecv{{\bf v}}

\def\hatk{{\bf \hat k}}

\def\l{{\ell}}
\def\eg{{\it e.g.~}}

\def\C{{\cal C}}
\def\P{{\cal P}}
\long\def\comment#1{}

\def\W2{{\cal W}}

\newcommand{\bi}{B_{l_1 l_2 l_3}}

\def\ben{\begin{enumerate}}
\def\een{\end{enumerate}}
\def\bi{\begin{itemize}}
\def\ei{\end{itemize}}
\def\be{\begin{equation}}
\def\ee{\end{equation}}
\def\bea{\begin{eqnarray}}
\def\eea{\end{eqnarray}}

\def\simlt{\mathrel{\mathpalette\fun <}}
\def\simgt{\mathrel{\mathpalette\fun >}}

\def\mpc{\,{\rm Mpc}}

\def\cmm2{{\,\rm cm^{-2}}}
\def\cm2{{\,{\rm cm}^2}}
\def\cmm3{{\,{\rm cm}^{-3}}}
\def\gcmm3{{\,{\rm g\,cm^{-3}}}}
\def\kms{\,{\rm km\,s^{-1}}}

\def\fun#1#2{\lower3.6pt\vbox{\baselineskip0pt\lineskip.9pt
  \ialign{$\mathsurround=0pt#1\hfil##\hfil$\crcr#2\crcr\sim\crcr}}}

\newcommand{\apjl}{{\it Astrophys. J. Lett.\,}}

\newcommand{\aap}{{\it A\&A\,}}

\begin{document}
\title{Probing the largest cosmological scales with the CMB-Velocity correlation}

\author{Pablo Fosalba}
\email{fosalba@ieec.uab.es}
\affiliation{Institut de Ci\`encies de l'Espai, IEEC-CSIC, Campus UAB,
Facultat de Ci\`encies, Torre C5 par-2,  Barcelona 08193, Spain}
\author{Olivier Dor\'e}
\email{olivier@cita.utoronto.ca}
\affiliation{CITA, University of Toronto, 60 St George Street,
  Toronto, ON M5S 3H8, Canada}
\received{\today}
%
\begin{abstract}
Cross-correlation between the CMB and large-scale structure is
a powerful probe of dark-energy and gravity on the largest physical scales.
We introduce a novel estimator, the CMB-velocity correlation, that has most of his power
on large scales and that, at low redshift, 
delivers up to factor of two higher signal-to-noise ratio 
than the recently detected CMB-dark matter density correlation 
expected from the Integrated Sachs-Wolfe effect.
We propose to use a combination of peculiar velocities measured from
supernovae type Ia and kinetic Sunyaev-Zeldovich cluster surveys to reveal this
signal and forecast dark-energy constraints that can be achieved with future surveys.
We stress that low redshift peculiar velocity measurements should be exploited with 
complementary deeper large-scale structure surveys for precision cosmology.
\end{abstract}
%
\maketitle

\renewcommand{\thefootnote}{\arabic{footnote}}
\setcounter{footnote}{0}

\section{Introduction}

The Integrated Sachs-Wolfe (ISW) effect carries information on 
the evolution of the universe at low and moderate redshifts ($z \simlt 3$) and thus is a  
powerful probe of dark-energy (DE) properties \citep{sachs67,crittenden96,bean04}. 
Since the ISW effect is most sensitive 
to gravity on the largest scales (hundreds of $\mpc$), 
it also offers a novel way to distinguish
dark-energy from modified gravity theories \citep{lue04,garriga04,song06}. 
Measuring the ISW signal that contributes to the total CMB temperature power spectrum 
is mainly limited by sampling variance errors on large scales,
that has traditionally challenged its detection. An alternative path to 
extracting the ISW effect is by 
cross-correlating CMB maps with tracers of the large-scale structure, such as galaxies.

Recently, first detections of the ISW effect from cross-correlation analyses of WMAP data
 with different galaxy surveys, and x-ray maps have been obtained
\citep{boughn04,nolta04,fosalba04,fosalba03,scranton03,afshordi04a,cabre06} .
Reported detections are at the $~2-4~\sigma$ level
and favor a DE dominated universe, independent of Supernovae type Ia (SNe hereafter). 
Future galaxy surveys (such as DES, Pan-STARRS, WFMOS) 
hold the promise of raising the significance of these detections by going deeper and wider
to optimally sample the ISW signal. Primary anisotropies of the CMB are the dominant 
source of noise in such correlation analyses
that seriously limits the ability with which this probe can constrain cosmology 
\citep{afshordi04b,cabre07}. 
In addition, CMB-galaxy correlations are affected by systematics
such as galaxy bias and shot noise. 
Moreover reconstructing the gravitational potential (and its time evolution) 
from the density field is a noisy process involving second order spatial derivatives.

Alternatively, velocity flow measurements from estimates 
of the luminosity distance and redshift are unbiased tracers of the gravitational potential, 
and the latter can be simply reconstructed through first spatial derivatives of the previous.
Also, these measurements are not affected by shot-noise unlike galaxy densities (specially in 
deep samples).
Peculiar velocities have more power on the largest (linear) scales than the dark matter density,
what follows from the continuity equation (see below), 
and thus provides a natural counterpart to the density field 
in reconstructing the gravitational potential.
However probing large-scale velocity flows is not an easy endeavor. 
Although galaxy surveys now sample large volumes, the intrinsic inaccuracy plaguing distance
estimations (20-25\%) limits the volume available for such measurements 
\cite{strauss95,hudson00,feldman06}. 
Nonetheless better distance indicators and other
probes of the velocity fields are already used in cosmology but
their power to reveal the ISW effect and the physics behind it has
been neglected so far: SNe currently yield distance measurements with
a 5\%-10\% error \citep{riess97,astier05} and kinetic Sunyaev-Zeldovich (kSZ,\cite{sunyaev80}) 
could in principle deliver
redshift independent velocity errors around 100 km.$s^{-1}$. These
surveys already allow to map the large-scale velocity field at low $z$ 
with a good agreement with other probes \cite{riess97,haugboelle06,benson03}. 

In this paper we propose to use the large-scale velocity flows from low-redshift SNe and 
kSZ cluster surveys to measure the ISW effect from the CMB-velocity correlation.
We propose to include velocity measurements at low redshift in combination with 
deeper large-scale structure probes in tomographic analyses
to better probe DE and gravity on the largest scales.
The complementarity of velocities with respect to densities comes from
the larger signal-to-noise of the CMB-velocity correlation estimator at low redshift
(up to a factor of 2 gain depending on cosmology),
as we will show in \S\ref{sec:prospects} below.
Unless otherwise stated, we will focus on 
flat  $\Lambda$CDM  models with $\Omega_{DE}=0.75$,
$\Omega_b=0.05$,$n_s=1$,$h=0.7$,$\sigma_8=0.9$, and
assume large-scale structure sources follow a density
distribution $dn/dz $$\propto z^2
\exp\left[-(z/(\sqrt 2 z_m))^{1.5}\right]$, with a width, 
$\sigma_z \simeq z_m/2$, where $z_m$ is the survey median redshift.

\begin{figure}[t]
\begin{center}
\includegraphics[width=0.45\textwidth,height=0.35\textwidth]{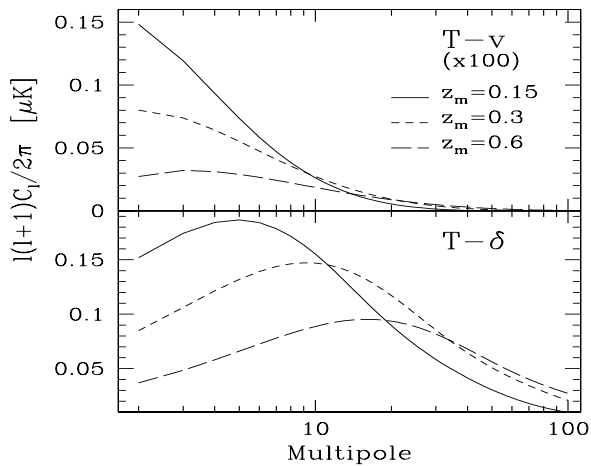}
\end{center}
\caption{Cross power spectra for the CMB temperature-velocity, $Tv$ (top), 
and the temperature-density, $T\delta$ (bottom), for different mean survey depths,
$z_m$. Most of the signal in the $Tv$ correlation comes from
larger scales (lower multipoles) than the $T\delta$. We amplify the
velocity correlation signal by 100 to match the ISW amplitude from 
the density correlation.}
\label{fig:cl_panel}
\end{figure}

\section{Modeling the Observables}
\label{sec:observables}

Dynamics of the fluctuations of the cosmological fields on the largest scales
can be accurately described by linear perturbation theory.
In this regime, the Poisson and continuity equation in
Fourier space are given by (Peebles 1980, \S 2)
$-{k^2}/{a^2}\Phi (k,t) = {3}/{2}H^2\Omega D\delta(k,t_0)$,
where we have used that $H^2 = (8\pi G/3) {\bar \rho}
/\Omega$, with $\bar \rho$ being the mean matter density, $a(t)$ is
the FRW scale factor, $k$ is the comoving wavenumber and $D(t)$
accounts for the linear growth of dark-matter density perturbations. 
The linear continuity equation reads,
$\vecv(k,t)= -i {\dot D} a\,\delta (k,t_0)/k\,\hatk$ 
where $\hatk=\veck/k$ and $\cdot$ denotes derivative with respect
to comoving time $t$. Combining both Fourier equations we get the linear evolution of
velocity modes as a function of the present day gravitational
potential (\eg \citep{dore03}),
$\vecv(k,t)= i\,k/H_0\,\Phi(k,t_0) g(t) \hatk$
where $\Phi(k,t) = D/a~\Phi(k,t_0)$ is the time evolution of gravitational potential modes and
$g(t)= 2a^2/3\Omega_0 \sqrt{\Omega_m/a^3+\Omega_\Lambda}~dD/da$.

Angular cross-power spectra between the CMB temperature anisotropy and the dark-matter
density $\delta$ and peculiar velocity $v$ fields 
can be derived from the corresponding angular 2-point correlation function 
$w_{XY}=\langle X(\vechr) Y(\vechr^\prime)\rangle = 
\sum (2\l+1)/4\pi~\C_{\l}^{XY} P_\l(\vechr\cdot\vechr^\prime)$,
and yields,
\be
\C_{\l}^{XY} = \frac{2}{\pi} \int k^2 dk\ P_{\Phi\Phi}(k) W_{\l}^X(k)W_\l^Y(k).
\label{eq:cps_def}
\ee
where 
the 3D power spectra is defined as, $(2\pi)^3\P_{XY}(k)\delta(\veck-\veck^\prime)=\langle \tilde
X(\veck)\tilde Y(\veck^\prime)\rangle$,
and the kernels $W_{\l}^X(k)$ that define how the signal from the $X$ field
spreads in comoving scales are given by (\citep{vishniac87})
\bea
W_{\l}^{T}(k) &=& \frac{2}{c^2}\int d\eta \dot{F}(\eta) j_{\l}(k\eta)\\
W_{\l}^{\delta} (k) & = & \frac{3k^2H_0^2\Omega_0}{2}\int d\eta F(\eta) \frac{dn(\eta)}{d\eta}j_\l(k\eta)\\
W_{\l}^v (k) & = & \frac{k}{H_0}\int d\eta ~\frac{dn(\eta)}{d\eta}x(\eta) j_\l^\prime(k\eta),
\eea
being $dn/d\eta$ the space density of sources and we denote
the radial velocity $v_{r} = |\vecv(\vecr)\cdot\vechr| \equiv v$. On
sufficiently large (linear) scales, gravitational instability generates
irrotational flows, thus we expect vanishing tangential component $v_\perp = 0$, 
(only the line-of-sight modes $k = k_\parallel$ are non-zero,\cite{kaiser84}) 
and we define $j^\prime(x) = dj(x)/dx$ with $j_{\l}(x)$ being 
the spherical Bessel function of order $\l$. In our computations, 
we shall use velocities in dimensionless units, by normalizing to the speed of light. 

\begin{figure}[t]
\begin{center}
\includegraphics[width=0.45\textwidth,height=0.35\textwidth]{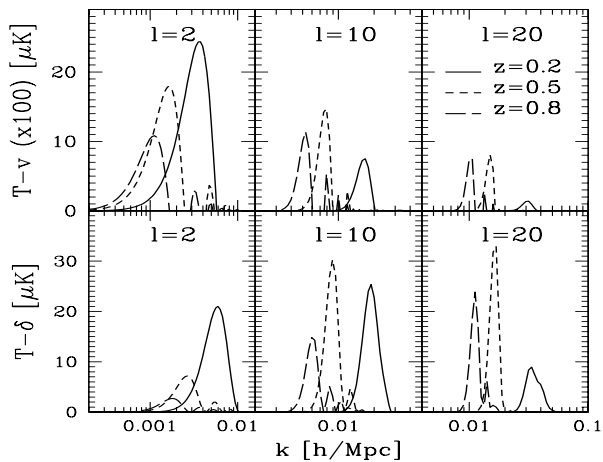}
\end{center}
\caption{Contribution to the ISW effect power spectra from different 
3D wavenumbers, $k$, and thin redshift $z$ shells, for various multipoles $\l$. 
As in Fig.~\ref{fig:cl_panel}, top panels display the $Tv$ kernel, 
amplified by a factor of 100,
whereas the bottom panels show the $T\delta$ kernel. Kernels are given by
$\l(\l+1)/2\pi$ times the integrand in eq.~(\ref{eq:cps_def}).} 

\label{fig:kern_kl}
\end{figure}

Fig.~\ref{fig:cl_panel} shows the CMB temperature-velocity, $Tv$, power 
is about 2 orders of magnitude smaller
than that of the temperature-density, $T\delta$, estimator. However, as we shall see below, the
noise contribution is more than 100 times smaller than for the
$T\delta$, what makes this signal easier to extract in cross-correlations analyses.
Note that most of the signal in the $Tv$ correlation
comes from larger scales (lower multipoles) than the $T\delta$ and therefore
very wide surveys are required to measure the expected correlation.
On the other hand, as illustrated by Fig.~\ref{fig:kern_kl}, the peak
amplitude of the $Tv$ kernels (i.e, integrand in eq.~(\ref{eq:cps_def})),
for relatively shallow surveys, $z \simlt 0.5$,
peaks at $\ell=2$, whereas for deeper surveys the signal is dominated by higher $\ell$'s.
Note that the CMB-velocity is mainly sensitive 
to modes at $k \sim 0.01-0.001 \mpc^{-1}$, i.e, almost one order of magnitude 
larger comoving scales than for $T\delta$.

\section{Observational prospects}
\label{sec:prospects}

Current probes of the velocity flow on large scales include galaxy peculiar velocities, 
SNe galaxy host peculiar velocities and the kinetic Sunyaev-Zel'dovich effect (kSZ).
The radial component of peculiar velocities of galaxies and
SNe, $v$ is determined by inferring
their distance, $d$, and then subtracting off the Hubble flow contribution to the
measured redshift, $z$, as $v=cz-H_0d$ \footnote{we ignored here the
velocity of the observer since it is well measured through the CMB
dipole}. 
The empirical correlation --between light-curve
shape and luminosity, and between color-curve shape and extinction--
used to infer the SNe luminosity distance currently yield a dispersion
in apparent magnitude, $m$, of $\sigma(m)=0.1-0.15$. Since $m$ is
related to the luminosity distance in Mpc, $d_L$, and the absolute magnitude, 
$M$, as  $m =$$ 5\log_{10}d_L^{}+M$, a dispersion of $\sigma(m)=0.1$ entails a redshift
independent distance measurements errors of 5\% when ignoring the
effect of the marginalization over the constant $M$. Therefore measuring 
the velocity of the host galaxy of one SNe can be as accurate as the 
velocity obtained from 25 galaxies using the $D-\sigma$
relation. In turn, the redshift for nearby galaxies (SNe host or not) 
is currently measured using narrow lines from the host galaxies with an uncertainty $\sigma(cz) =
30\mbox{km.s}^{-1}$ that is dominated by statistical errors. 
As such for a Hubble constant of $h=0.72$, errors
in redshift measurements will be subdominant as compared to errors in $H_0d$ for
$z>$0.02 (0.004) if we use SNe (galaxy) based distance estimators, respectively.  
Since galaxy peculiar radial velocities typically have a rms of 300
$\mbox{km.s}^{-1}$ we reach a signal-to-noise ratio $S/N \simeq 1$ per source at
$z\simeq 0.02 (0.004)$. If we were to volume-average the distance
measurements coming from SNe observations over samples at fixed z, the
systematic error limit would be $\sigma(m)=0.02$ \cite{kim04}. 
Assuming that we had enough SNe to reach this limit
within boxes of depth $\sigma(cz) =30\mbox{km.s}^{-1}$, then 
one could get a 1\% distance measurement and the S/N per source
would reach 1 at $z\simeq 0.1$.

On the other hand, the kSZ determined peculiar velocities of \emph{galaxy clusters} do not
rely on a distance determination, and associated
$z$-independent errors of $\sim 100\kms$ may be achievable
\citep{nagai03,holder04}. However the signal is limited to $z \simgt 0.1$ in
order for the primordial CMB not to be a dominant source of
confusion. Its power as a tracer of the large-scale gravitational potential 
has already been studied in \cite{dore03} and as a DE probe
in conjunction with other density tracers in \cite{dore04,dedeo05}.

The nature of the signal we are investigating demands wide but 
rather shallow surveys in comparison to other probes.  
To illustrate the detectability of this new correlation, we will consider two ideal full-sky
surveys, with a distribution of sources with median redshift 
$z_m=0.15$ or $z_m=0.3$, and study their performance in 
ISW detection and DE constraints as compared to the $T\delta$ correlation. 
In particular, we propose to combine full-sky peculiar velocity surveys from SNe
at  $z \simlt 0.1$ and, complementarily, kSZ cluster surveys for
sources at $z \simgt 0.1$.
\begin{figure}[t]
\begin{center}
\includegraphics[width=0.45\textwidth,height=0.35\textwidth]{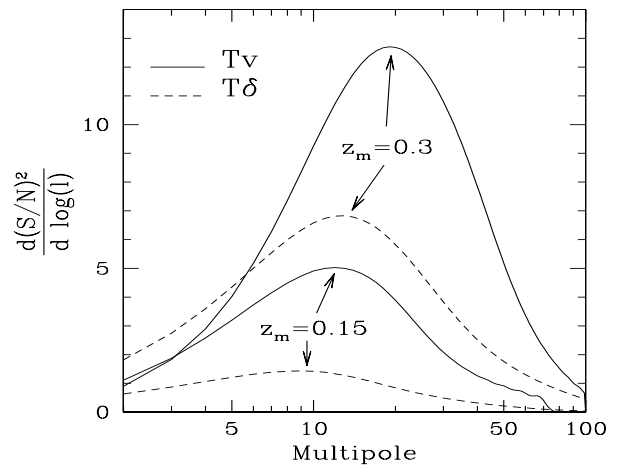}
\end{center}
\caption{Contribution of multipoles to the signal-to-noise ratio S/N 
for $Tv$ (solid) and $T\delta$ (dashed) for full-sky surveys with different depths, 
for our baseline $\Lambda$CDM model. 
Total (S/N)$^2$ is given by the area under the curves.
Velocities give a S/N in cross-correlation with the temperature that is 
systematically larger than that for densities, $T\delta$, 
the gain factor depending on survey depth and cosmology. 
Note that S/N scales as $\sqrt{f_{sky}}$, 
where $f_{sky}$ is the fraction of sky covered by the survey.}
\label{fig:s2n_panel}
\end{figure}
We define the cross-correlation signal-to-noise ratio S/N as
\be
(S/N)^2=\sum_{\ell}(2\ell+1)f_{sky}~\frac{(C_\ell^{TX})^2}{(C_\ell^{TX})^2+C_\ell^{TT}C_\ell^{XX}}
\label{eq:sn_def}
\ee
where $X = v$ or $\delta$. 
Fig.~\ref{fig:s2n_panel} shows the contribution of different multipoles to the 
S/N depending on survey depth, for a $\Lambda$CDM model
with $\Omega_{DE}=0.75$ that will be our baseline. 
The area under the curves give the total (S/N)$^2$.
We find that for low z, 
$Tv$ peaks at similar (albeit slightly larger) $\ell$'s than $T\delta$, i.e, $\ell\sim 10-20$.
For our baseline cosmology one gets, for the $z_m=0.3$ survey, 
(S/N)$_{Tv} = 5$, that is 25\% 
larger than the corresponding 
significance for the ISW detection from the $T\delta$ correlation.
The shallower survey, $z_m=0.15$, leads to more moderate significance, (S/N)$_{Tv} = 3.2$. 
However, this is about factor of $2$ larger than that for
$T\delta$. This comes from the fact that the lowest redshift sources give the dominant 
contribution to the overall S/N.
In other words, as shown by Fig.~\ref{fig:kern_kl}, 
progressively deeper velocity surveys are less optimal for
ISW measurements since the $Tv$ signal quickly drops with redshift unlike $T\delta$.
In general, the relative gain in S/N when using velocities as compared to densities 
can be understood
by computing the ratio of S/N for $Tv$ with respect to $T\delta$ for a given multipole $\ell$,
that scales as $(C_\ell^{Tv}/C_\ell^{T\delta})\sqrt{C_\ell^{\delta\delta}/C_\ell^{vv}}$. 
Although at low multipoles the signal is about two orders of magnitude smaller for velocities, 
$(C_\ell^{Tv}/C_\ell^{T\delta}) \sim 0.01$ 
(see Fig.\ref{fig:cl_panel}), the corresponding 
ratio of noise terms arising from the auto-correlations 
$C_\ell^{\delta\delta}/C_\ell^{vv} \simgt 10^4$, 
what makes the relative S/N exceed unity.
We note that this $Tv$ significance gain over $T\delta$
increases for more strongly DE dominated cosmologies.
In particular, for a shallow survey ($z_m=0.15$), velocities
can measure the cosmic low multipoles with a significance $\simgt 2$ 
times larger than dark-matter density tracers, such as galaxies, in 
cross-correlation with the CMB.

\begin{figure}[t]
\begin{center}
\includegraphics[width=0.45\textwidth,height=0.4\textwidth]{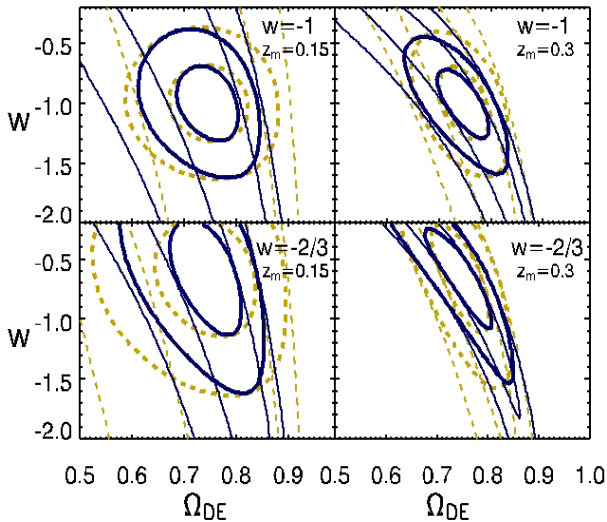}
\end{center}
\caption{Constraints on the DE parameters 
from the cross-correlation of the CMB with shallow full-sky surveys. 
Contours show 68\% (inner line) and 95\% (outer line) confidence levels for 1 DOF. 
Solid (dashed) lines show parameter errors from the $Tv$ ($T\delta$) 
correlation, whereas thick (thin) lines correspond to constraints 
with (without) PLANCK priors.
Left (right) panel shows the case for surveys with median redshift 
$z_m=0.15$ ($z_m=0.3$) and top (bottom) panels display two different choices
of the DE equation of state: $w=-1$ ($\Lambda$CDM) and $-2/3$ (quintessence).}
\label{fig:cosmo_panel2}
\end{figure}

\section{Cosmological impact and conclusions}
\label{sec:discuss}

In the context of flat CDM models, an increased significance 
in the ISW measurement from the $Tv$ correlation directly
translates into tighter DE constraints. 
Fig.~\ref{fig:cosmo_panel2} displays the expected parameter constraints
that can be achieved for two shallow full-sky surveys. For comparison purposes, 
we show constraints from $Tv$ (solid lines) and the usual $T\delta$ (dashed) correlations.
In general, the $Tv$ estimator yields better DE constraints than the $T\delta$
irrespective of the baseline cosmology used or survey depth, provided $z_m \simlt 0.3$,
beyond which the usual $T\delta$ cross-correlation outperforms $Tv$.
A shallow velocity survey with $z_m=0.15$ can deliver a (symmetrized) error $\sigma(\Omega_{DE})=0.065$
for our baseline $\Lambda$CDM cosmology, that is almost a factor of 2 better accuracy than 
what densities can deliver (see e.g, lower left panel
in Fig~\ref{fig:cosmo_panel2}). These constraints are comparable to what 
is expected for $\sigma(\Omega_{DE})$ from PLANCK\footnote{we use Fisher matrix priors 
from PLANCK temperature and polarization data, as given by \citep{hu02}, i.e, we assume
$\sigma(\Omega_{DE})=0.1$, $\sigma(w)=0.32$ for $\Lambda$CDM with $\Omega_{DE}=0.75$, and 
$\sigma(\Omega_{DE})=0.2$, $\sigma(w)=0.5$ for a quintessence model with $w=-2/3$ and $\Omega_{DE}=0.75$.}
(see comparison between thin and thick solid lines in the left panels), 
although such an outstanding CMB dataset will
largely help breaking the degeneracy with $w$. For the deeper survey, $z_m=0.3$, 
velocities yield a factor of 2 better measurement of the DE density $\sigma(\Omega_{DE})=0.03$ than
the shallower survey (solid contours in right panels as compared to same lines in left panels),
but this is comparable to what a density tracer would deliver (dashed contours in corresponding panels).
Velocities can also help breaking the degeneracy with $w$, 
but, as expected from a low redshift survey, to a much lower extent than PLANCK.
Note however how the DE errors depend on the baseline cosmology used: 
$Tv$ measurements can provide significanlty sharper constraints on quintessence models 
than on $\Lambda$CDM and yield better measurements of $\Omega_{DE}$ than PLANCK.
This is shown in the lower panels of Fig~\ref{fig:cosmo_panel2} where errors from the 
$Tv$ correlation alone (thin solid lines) do not improve when adding PLANCK priors (thick solid lines).

In practice several factors could degrade these optimal expectations. 
In cross-correlations S/N $\propto \sigma_8$ and thus
current estimates of $\sigma_8\simeq 0.8$ would imply a 10\% weaker detection 
than we estimated from our baseline model. 
Lensing could introduce noise in $Tv$ as we go deeper in redshift
although its large scale contribution is small. More importantly, we
neglected so far the fact that we deal with a discrete sampling of the
underlying velocity field. Note that for velocities no shot noise affects their measurement,
unlike the case for densities. However, if we consider the case for a
$z_m=0.3$ survey, 90\% of sources lie below $z=0.8$. 
Since there are around $10^5$ clusters with masses greater than $10^{14}$ h$^{-1}$ M$_{\odot}$
in this volume, a full-sky cluster kSZ survey
up to $z=0.8$ would provide a good sampling of the
field for modes $k\simlt 0.1$ Mpc$^{-1}$ which is sufficient for the
CMB-velocity signal according to Fig.~\ref{fig:kern_kl}. 

In conclusion, the use of wide and shallow velocity surveys  can 
substantially increase our ability to extract the ISW signal from a
cross-correlation with the CMB and thus 
help other cosmological probes to distinguish gravity from alternatives.
Peculiar velocities are highly correlated with other large-scale structure tracers
such as the galaxy density distribution but probe larger physical scales.
In addition, the newly proposed correlation is a priori less affected by systematics
that plague the temperature-galaxy correlation such as galaxy bias and shot noise.
In particular, combining shallow velocity surveys with complementary data from 
planned deep surveys (such as LSST, DES, Pan-STARRS, WFMOS)
can fully exploit the ability of tomographic analyses of future surveys 
to reveal the nature of dark-energy and gravity on the largest
physical scales \citep{hu02,kesden03}. 
Furthermore, the temperature-velocity correlation 
is a more powerful probe of the CMB lowest multipoles 
than the standard correlation with dark-matter density tracers,
such as galaxies, and hence it could shed new light on inflation and the 
physics of the primordial universe.

\acknowledgments
We are grateful to Niayesh Afshordi, Pierre Astier, Enrique Gazta\~{n}aga, Joe Hennawi, Wayne
Hu, Mike Hudson, Arthur Kosowsky, Robert Lupton, Ram\'on Miquel, Jordi Miralda-Escud\'e, Pat McDonald, Bohdan Paczynski, David Schlegel, David Spergel, and Martin White for useful comments.
PF acknowledges support from the Spanish MEC through a Ramon y Cajal fellowship, and grants 
AYA2006-06341, 2005SGR-00728.

\appendix

\bibliography{vl}

\begin{thebibliography}{33}
\expandafter\ifx\csname natexlab\endcsname\relax\def\natexlab#1{#1}\fi
\expandafter\ifx\csname bibnamefont\endcsname\relax
  \def\bibnamefont#1{#1}\fi
\expandafter\ifx\csname bibfnamefont\endcsname\relax
  \def\bibfnamefont#1{#1}\fi
\expandafter\ifx\csname citenamefont\endcsname\relax
  \def\citenamefont#1{#1}\fi
\expandafter\ifx\csname url\endcsname\relax
  \def\url#1{\texttt{#1}}\fi
\expandafter\ifx\csname urlprefix\endcsname\relax\def\urlprefix{URL }\fi
\providecommand{\bibinfo}[2]{#2}
\providecommand{\eprint}[2][]{\url{#2}}

\bibitem[{\citenamefont{{Sachs} and {Wolfe}}(1967)}]{sachs67}
\bibinfo{author}{\bibfnamefont{R.~K.} \bibnamefont{{Sachs}}} \bibnamefont{and}
  \bibinfo{author}{\bibfnamefont{A.~M.} \bibnamefont{{Wolfe}}},
  \bibinfo{journal}{\apj} \textbf{\bibinfo{volume}{147}}, \bibinfo{pages}{73}
  (\bibinfo{year}{1967}).

\bibitem[{\citenamefont{{Crittenden} and {Turok}}(1996)}]{crittenden96}
\bibinfo{author}{\bibfnamefont{R.~G.} \bibnamefont{{Crittenden}}}
  \bibnamefont{and} \bibinfo{author}{\bibfnamefont{N.}~\bibnamefont{{Turok}}},
  \bibinfo{journal}{Physical Review Letters} \textbf{\bibinfo{volume}{76}},
  \bibinfo{pages}{575} (\bibinfo{year}{1996}).

\bibitem[{\citenamefont{{Bean} and {Dor{\'e}}}(2004)}]{bean04}
\bibinfo{author}{\bibfnamefont{R.}~\bibnamefont{{Bean}}} \bibnamefont{and}
  \bibinfo{author}{\bibfnamefont{O.}~\bibnamefont{{Dor{\'e}}}},
  \bibinfo{journal}{\prd} \textbf{\bibinfo{volume}{69}},
  \bibinfo{pages}{083503} (\bibinfo{year}{2004}).

\bibitem[{\citenamefont{{Lue}~\etal}(2004)}]{lue04}
\bibinfo{author}{\bibfnamefont{A.}~\bibnamefont{{Lue}~\etal}},
  \bibinfo{journal}{\prd} \textbf{\bibinfo{volume}{69}},
  \bibinfo{pages}{044005} (\bibinfo{year}{2004}).

\bibitem[{\citenamefont{{Garriga}~\etal}(2004)}]{garriga04}
\bibinfo{author}{\bibfnamefont{J.}~\bibnamefont{{Garriga}~\etal}},
  \bibinfo{journal}{\prd} \textbf{\bibinfo{volume}{69}},
  \bibinfo{pages}{063511} (\bibinfo{year}{2004}).

\bibitem[{\citenamefont{{Song} et~al.}(2006)\citenamefont{{Song}, {Sawicki},
  and {Hu}}}]{song06}
\bibinfo{author}{\bibfnamefont{Y.-S.} \bibnamefont{{Song}}},
  \bibinfo{author}{\bibfnamefont{I.}~\bibnamefont{{Sawicki}}},
  \bibnamefont{and} \bibinfo{author}{\bibfnamefont{W.}~\bibnamefont{{Hu}}},
  \bibinfo{journal}{ArXiv Astrophysics e-prints}  (\bibinfo{year}{2006}),
  \eprint{astro-ph/0606286}.

\bibitem[{\citenamefont{{Boughn} and {Crittenden}}(2004)}]{boughn04}
\bibinfo{author}{\bibfnamefont{S.}~\bibnamefont{{Boughn}}} \bibnamefont{and}
  \bibinfo{author}{\bibfnamefont{R.}~\bibnamefont{{Crittenden}}},
  \bibinfo{journal}{\nat} \textbf{\bibinfo{volume}{427}}, \bibinfo{pages}{45}
  (\bibinfo{year}{2004}).

\bibitem[{\citenamefont{{Nolta}~\etal}(2004)}]{nolta04}
\bibinfo{author}{\bibfnamefont{M.~R.} \bibnamefont{{Nolta}~\etal}},
  \bibinfo{journal}{\apj} \textbf{\bibinfo{volume}{608}}, \bibinfo{pages}{10}
  (\bibinfo{year}{2004}).

\bibitem[{\citenamefont{{Fosalba} and {Gazta{\~n}aga}}(2004)}]{fosalba04}
\bibinfo{author}{\bibfnamefont{P.}~\bibnamefont{{Fosalba}}} \bibnamefont{and}
  \bibinfo{author}{\bibfnamefont{E.}~\bibnamefont{{Gazta{\~n}aga}}},
  \bibinfo{journal}{\mnras} \textbf{\bibinfo{volume}{350}},
  \bibinfo{pages}{L37} (\bibinfo{year}{2004}).

\bibitem[{\citenamefont{{Fosalba}~\etal}(2003)}]{fosalba03}
\bibinfo{author}{\bibfnamefont{P.}~\bibnamefont{{Fosalba}~\etal}},
  \bibinfo{journal}{\apjl} \textbf{\bibinfo{volume}{597}}, \bibinfo{pages}{L89}
  (\bibinfo{year}{2003}).

\bibitem[{\citenamefont{{Scranton}~\etal}(2003)}]{scranton03}
\bibinfo{author}{\bibfnamefont{R.}~\bibnamefont{{Scranton}~\etal}},
  \bibinfo{journal}{ArXiv Astrophysics e-prints}  (\bibinfo{year}{2003}),
  \eprint{astro-ph/0307335}.

\bibitem[{\citenamefont{{Afshordi} et~al.}(2004)\citenamefont{{Afshordi},
  {Loh}, and {Strauss}}}]{afshordi04a}
\bibinfo{author}{\bibfnamefont{N.}~\bibnamefont{{Afshordi}}},
  \bibinfo{author}{\bibfnamefont{Y.-S.} \bibnamefont{{Loh}}}, \bibnamefont{and}
  \bibinfo{author}{\bibfnamefont{M.~A.} \bibnamefont{{Strauss}}},
  \bibinfo{journal}{\prd} \textbf{\bibinfo{volume}{69}},
  \bibinfo{pages}{083524} (\bibinfo{year}{2004}).

\bibitem[{\citenamefont{{Cabre}~\etal}(2006)}]{cabre06}
\bibinfo{author}{\bibfnamefont{A.}~\bibnamefont{{Cabre}~\etal}},
  \bibinfo{journal}{ArXiv Astrophysics e-prints}  (\bibinfo{year}{2006}),
  \eprint{astro-ph/0603690}.

\bibitem[{\citenamefont{{Afshordi}}(2004)}]{afshordi04b}
\bibinfo{author}{\bibfnamefont{N.}~\bibnamefont{{Afshordi}}},
  \bibinfo{journal}{\prd} \textbf{\bibinfo{volume}{70}},
  \bibinfo{pages}{083536} (\bibinfo{year}{2004}).

\bibitem[{\citenamefont{{Cabre} et~al.}(2007)\citenamefont{{Cabre}, {Fosalba},
  {Gaztanaga}, and {Manera}}}]{cabre07}
\bibinfo{author}{\bibfnamefont{A.}~\bibnamefont{{Cabre}}},
  \bibinfo{author}{\bibfnamefont{P.}~\bibnamefont{{Fosalba}}},
  \bibinfo{author}{\bibfnamefont{E.}~\bibnamefont{{Gaztanaga}}},
  \bibnamefont{and} \bibinfo{author}{\bibfnamefont{M.}~\bibnamefont{{Manera}}},
  \bibinfo{journal}{ArXiv Astrophysics e-prints}  (\bibinfo{year}{2007}),
  \eprint{astro-ph/0701393}.

\bibitem[{\citenamefont{{Strauss} and {Willick}}(1995)}]{strauss95}
\bibinfo{author}{\bibfnamefont{M.~A.} \bibnamefont{{Strauss}}}
  \bibnamefont{and} \bibinfo{author}{\bibfnamefont{J.~A.}
  \bibnamefont{{Willick}}}, \bibinfo{journal}{\physrep}
  \textbf{\bibinfo{volume}{261}}, \bibinfo{pages}{271} (\bibinfo{year}{1995}).

\bibitem[{\citenamefont{{Hudson}~\etal}(2000)}]{hudson00}
\bibinfo{author}{\bibfnamefont{M.~J.} \bibnamefont{{Hudson}~\etal}}, in
  \emph{\bibinfo{booktitle}{ASP Conf. Ser. 201: Cosmic Flows Workshop}}, edited
  by \bibinfo{editor}{\bibfnamefont{S.}~\bibnamefont{{Courteau}}}
  \bibnamefont{and} \bibinfo{editor}{\bibfnamefont{J.}~\bibnamefont{{Willick}}}
  (\bibinfo{year}{2000}), pp. \bibinfo{pages}{159--+}.

\bibitem[{\citenamefont{{Sarkar}~\etal}(2006)}]{feldman06}
\bibinfo{author}{\bibfnamefont{D.}~\bibnamefont{{Sarkar}~\etal}},
  \bibinfo{journal}{ArXiv Astrophysics e-prints}  (\bibinfo{year}{2006}),
  \eprint{astro-ph/0607426}.

\bibitem[{\citenamefont{{Riess}~\etal}(1997)}]{riess97}
\bibinfo{author}{\bibfnamefont{A.~G.} \bibnamefont{{Riess}~\etal}},
  \bibinfo{journal}{\apjl} \textbf{\bibinfo{volume}{488}}, \bibinfo{pages}{L1+}
  (\bibinfo{year}{1997}).

\bibitem[{\citenamefont{{Astier}~\etal}(2006)}]{astier05}
\bibinfo{author}{\bibfnamefont{P.}~\bibnamefont{{Astier}~\etal}},
  \bibinfo{journal}{\aap} \textbf{\bibinfo{volume}{447}}, \bibinfo{pages}{31}
  (\bibinfo{year}{2006}).

\bibitem[{\citenamefont{{Sunyaev} and {Zeldovich}}(1980)}]{sunyaev80}
\bibinfo{author}{\bibfnamefont{R.~A.} \bibnamefont{{Sunyaev}}}
  \bibnamefont{and} \bibinfo{author}{\bibfnamefont{I.~B.}
  \bibnamefont{{Zeldovich}}}, \bibinfo{journal}{\mnras}
  \textbf{\bibinfo{volume}{190}}, \bibinfo{pages}{413} (\bibinfo{year}{1980}).

\bibitem[{\citenamefont{{Haugboelle}~\etal}(2006)}]{haugboelle06}
\bibinfo{author}{\bibfnamefont{T.}~\bibnamefont{{Haugboelle}~\etal}},
  \bibinfo{journal}{ArXiv Astrophysics e-prints}  (\bibinfo{year}{2006}),
  \eprint{astro-ph/0612137}.

\bibitem[{\citenamefont{{Benson}~\etal}(2003)}]{benson03}
\bibinfo{author}{\bibfnamefont{B.~A.} \bibnamefont{{Benson}~\etal}},
  \bibinfo{journal}{\apj} \textbf{\bibinfo{volume}{592}}, \bibinfo{pages}{674}
  (\bibinfo{year}{2003}).

\bibitem[{\citenamefont{{Dor{\'e}}~\etal}(2003)}]{dore03}
\bibinfo{author}{\bibfnamefont{O.}~\bibnamefont{{Dor{\'e}}~\etal}},
  \bibinfo{journal}{\apjl} \textbf{\bibinfo{volume}{585}}, \bibinfo{pages}{L81}
  (\bibinfo{year}{2003}).

\bibitem[{\citenamefont{{Vishniac}}(1987)}]{vishniac87}
\bibinfo{author}{\bibfnamefont{E.~T.} \bibnamefont{{Vishniac}}},
  \bibinfo{journal}{\apj} \textbf{\bibinfo{volume}{322}}, \bibinfo{pages}{597}
  (\bibinfo{year}{1987}).

\bibitem[{\citenamefont{{Kaiser}}(1984)}]{kaiser84}
\bibinfo{author}{\bibfnamefont{N.}~\bibnamefont{{Kaiser}}},
  \bibinfo{journal}{\apj} \textbf{\bibinfo{volume}{282}}, \bibinfo{pages}{374}
  (\bibinfo{year}{1984}).

\bibitem[{\citenamefont{{Kim}~\etal}(2004)}]{kim04}
\bibinfo{author}{\bibfnamefont{A.~G.} \bibnamefont{{Kim}~\etal}},
  \bibinfo{journal}{\mnras} \textbf{\bibinfo{volume}{347}},
  \bibinfo{pages}{909} (\bibinfo{year}{2004}).

\bibitem[{\citenamefont{{Nagai} et~al.}(2003)\citenamefont{{Nagai}, {Kravtsov},
  and {Kosowsky}}}]{nagai03}
\bibinfo{author}{\bibfnamefont{D.}~\bibnamefont{{Nagai}}},
  \bibinfo{author}{\bibfnamefont{A.~V.} \bibnamefont{{Kravtsov}}},
  \bibnamefont{and}
  \bibinfo{author}{\bibfnamefont{A.}~\bibnamefont{{Kosowsky}}},
  \bibinfo{journal}{\apj} \textbf{\bibinfo{volume}{587}}, \bibinfo{pages}{524}
  (\bibinfo{year}{2003}).

\bibitem[{\citenamefont{{Holder}}(2004)}]{holder04}
\bibinfo{author}{\bibfnamefont{G.~P.} \bibnamefont{{Holder}}},
  \bibinfo{journal}{\apj} \textbf{\bibinfo{volume}{602}}, \bibinfo{pages}{18}
  (\bibinfo{year}{2004}).

\bibitem[{\citenamefont{{Dor{\'e}} et~al.}(2004)\citenamefont{{Dor{\'e}},
  {Hennawi}, and {Spergel}}}]{dore04}
\bibinfo{author}{\bibfnamefont{O.}~\bibnamefont{{Dor{\'e}}}},
  \bibinfo{author}{\bibfnamefont{J.~F.} \bibnamefont{{Hennawi}}},
  \bibnamefont{and} \bibinfo{author}{\bibfnamefont{D.~N.}
  \bibnamefont{{Spergel}}}, \bibinfo{journal}{\apj}
  \textbf{\bibinfo{volume}{606}}, \bibinfo{pages}{46} (\bibinfo{year}{2004}).

\bibitem[{\citenamefont{{DeDeo}~\etal}(2005)}]{dedeo05}
\bibinfo{author}{\bibfnamefont{S.}~\bibnamefont{{DeDeo}~\etal}},
  \bibinfo{journal}{ArXiv Astrophysics e-prints}  (\bibinfo{year}{2005}),
  \eprint{astro-ph/0511060}.

\bibitem[{\citenamefont{{Hu}}(2002)}]{hu02}
\bibinfo{author}{\bibfnamefont{W.}~\bibnamefont{{Hu}}}, \bibinfo{journal}{\prd}
  \textbf{\bibinfo{volume}{65}}, \bibinfo{pages}{023003}
  (\bibinfo{year}{2002}), \eprint{astro-ph/0108090}.

\bibitem[{\citenamefont{{Kesden} et~al.}(2003)\citenamefont{{Kesden},
  {Kamionkowski}, and {Cooray}}}]{kesden03}
\bibinfo{author}{\bibfnamefont{M.}~\bibnamefont{{Kesden}}},
  \bibinfo{author}{\bibfnamefont{M.}~\bibnamefont{{Kamionkowski}}},
  \bibnamefont{and} \bibinfo{author}{\bibfnamefont{A.}~\bibnamefont{{Cooray}}},
  \bibinfo{journal}{Physical Review Letters} \textbf{\bibinfo{volume}{91}},
  \bibinfo{pages}{221302} (\bibinfo{year}{2003}).

\end{thebibliography}


\end{document}